\documentclass[conference]{IEEEtran}
\IEEEoverridecommandlockouts
\usepackage{cite}
\usepackage{amsmath,amssymb,amsfonts}
\usepackage{algorithmic}
\usepackage{graphicx}
\usepackage{textcomp}
\usepackage{xcolor}

\newif\iffinal

\def\BibTeX{{\rm B\kern-.05em{\sc i\kern-.025em b}\kern-.08em
    T\kern-.1667em\lower.7ex\hbox{E}\kern-.125emX}}
\begin{document}
\bstctlcite{IEEEexample:BSTcontrol}

\title{The Manufacturing Data and Machine Learning Platform: Enabling Real-time Monitoring and Control of Scientific Experiments via IoT}

\author{
    \IEEEauthorblockN{Jakob R. Elias\IEEEauthorrefmark{1}, Ryan Chard\IEEEauthorrefmark{2}, Joseph A. Libera\IEEEauthorrefmark{3}, Ian Foster\IEEEauthorrefmark{2}\IEEEauthorrefmark{4}, Santanu Chaudhuri\IEEEauthorrefmark{1}\IEEEauthorrefmark{5}}
    \IEEEauthorblockA{\IEEEauthorrefmark{1}Energy and Global Security, Argonne National Laboratory:
    \{jelias, schaudhuri\}@anl.gov}
    \IEEEauthorblockA{\IEEEauthorrefmark{2}Data Science and Learning Division, Argonne National Laboratory: 
    \{rchard, foster\}@anl.gov}
     \IEEEauthorblockA{\IEEEauthorrefmark{3}Energy Systems Division, Argonne National Laboratory:
    jlibera@anl.gov}
     \IEEEauthorblockA{\IEEEauthorrefmark{4}Department of Computer Science, University of Chicago}
     \IEEEauthorblockA{\IEEEauthorrefmark{5}University of Illinois at Chicago: santc@uic.edu}
}

\maketitle

\begin{abstract}

IoT devices and sensor networks present new opportunities for measuring,
monitoring, and guiding scientific experiments. Sensors, cameras, and instruments can be combined to provide previously unachievable insights into the state of ongoing experiments. However, IoT devices can vary greatly in the type,
volume, and velocity of data they generate, making it challenging to fully realize this potential. Indeed, synergizing diverse IoT data streams in near-real time can require the use of machine learning (ML). In addition, new tools and technologies are required to facilitate the collection, aggregation, and manipulation of sensor data in order to simplify the application of ML models and in turn, fully realize the utility of IoT devices in laboratories. Here we will demonstrate how the use of the Argonne-developed Manufacturing Data and Machine Learning (MDML) platform 
can analyze and use IoT devices in a manufacturing experiment. MDML is designed to standardize the research and operational environment for advanced data analytics and AI-enabled automated process optimization by providing the infrastructure to integrate AI in cyber-physical systems for in situ analysis. We will show that MDML is capable of processing diverse IoT data streams, using multiple computing resources, and integrating ML models to guide an experiment.
\end{abstract}


\section{Introduction}

IoT devices~\cite{singh2014survey} are
becoming increasingly common in research laboratories~\cite{shumate2018iot},
where they promise to transform experimentation by enabling the concurrent measurement of many experimental characteristics in real time. 
Integrating IoT data sources presents exciting new opportunities to attain previously unachievable insights into the state of an ongoing experiment, and to steer and optimize instruments in real time. However, these opportunities also come with new challenges. The many different sensor types results in data streams with widely disparate rates, volumes, and velocities. These streams must be dynamically aggregated, transformed, and analyzed in order to perform in situ analysis---tasks that are infeasible for human operators. 

Machine Learning (ML) has been shown to be an effective tool for analyzing big IoT data streams~\cite{marjani2017big}. However, deploying and using ML models 
can require the use of specialized methods, software, and environments~\cite{dlhub}.
In addition, the deluge of data generated by IoT devices and the addition of new sensors can quickly exceed the processing capabilities of computing resources colocated with experimental apparatus. 
Achieving near-real-time analysis to enable online feedback can require the use of remote high performance computing (HPC) systems and specialized ML accelerators. 

The Manufacturing Data and Machine Learning (MDML) platform standardizes the research and operational environment for advanced data analytics to enable automated, ML-driven optimization. MDML is designed 
to support in situ measurements for accelerating scalable materials manufacturing
and to enable the integration of ML and HPC resources into the experimental process. 
MDML enables users to construct rich, data-oriented analysis pipelines that span disparate computational environments, from laptops and local servers to supercomputers and clouds. Finally, MDML leverages industry standard visualization and monitoring tools to create dynamic interfaces to experimental facilities and their processing pipelines.

We will demonstrate MDML and its application to the Manufacturing Engineering Research Facility's (MERF) combustion synthesis research project, in which ML-guided steering is used to enable the high-throughput manufacturing of nanomaterials. In particular, we will showcase data being streamed from multiple combustion chamber sensors into MDML, and then processed in near real time
to steer the experimental configuration. The demonstration will make use of local MERF servers and Argonne National Laboratory's Theta supercomputer, while visualizing the state of the experiment and data processing in a user-friendly interface.

\section{The MDML Platform}

The design of the MDML platform is motivated by the unique needs and challenging demands of MERF's scale-up research and development. MDML is designed to enable scientists to dynamically
integrate scientific IoT device data during high throughput experiments 
to guide and automatically 
optimize experiments. 
As such, MDML simplifies the aggregation, analysis, and application of ML 
to IoT data streams with almost arbitrary data rates and volumes

\begin{figure}[h]
 \centering
 \includegraphics[width=\columnwidth,trim=0in 0.9in 1.7in 0in,clip]{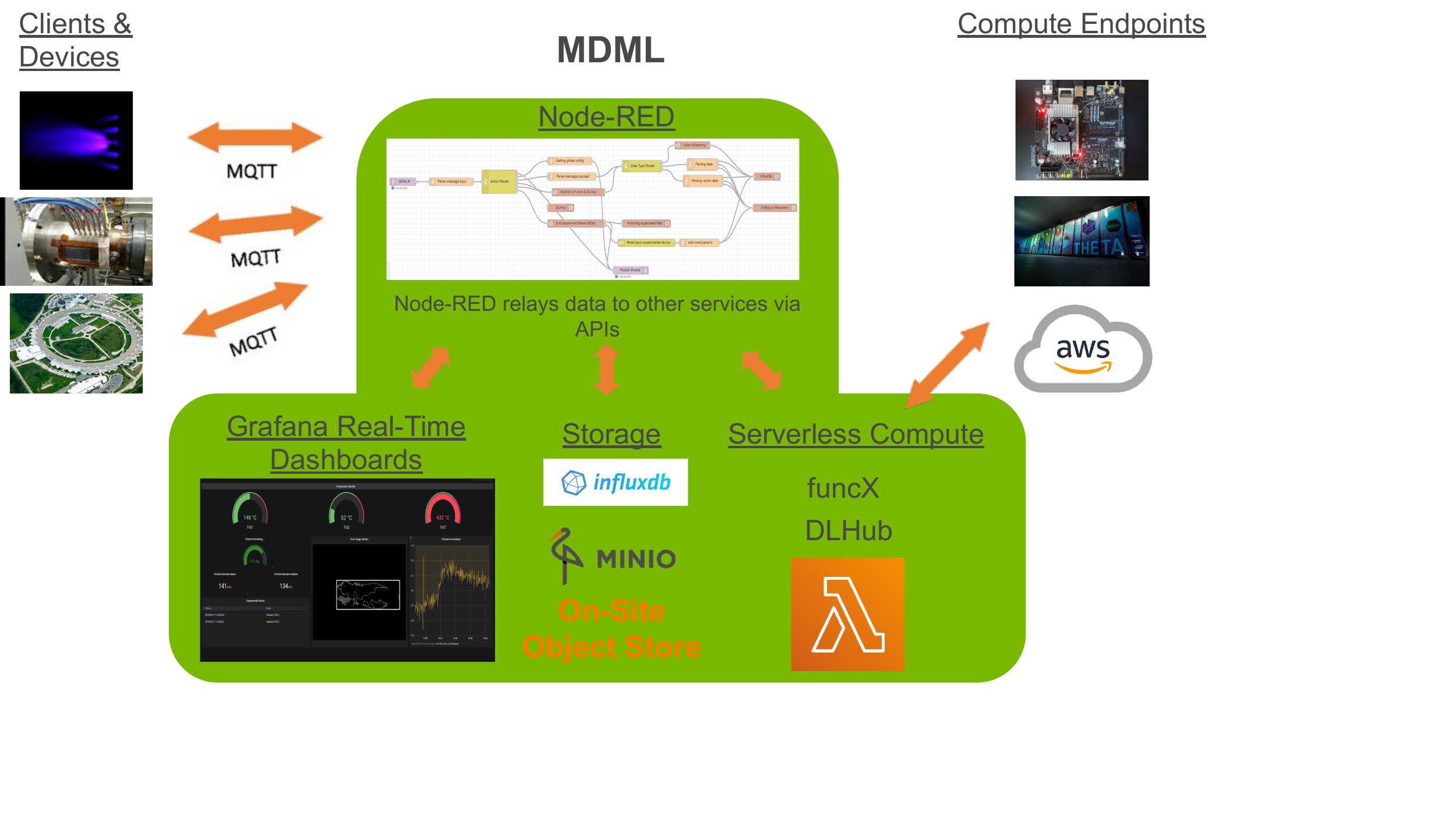}
 \caption{MDML architecture. Client devices (left) transmit data via MQTT to the MDML platform for storage and analysis. MDML uses funcX to offload computation to different computing resources.}
 \label{fig:mdml}
\end{figure}

The MDML platform, depicted in \figurename~\ref{fig:mdml}, is deployed at Argonne National Laboratory. 
MDML uses IBM's Node-RED platform~\cite{nodered}, designed for connected IoT assets, to integrate in situ measurements
from the experimental stations at the MERF. 
MQ Telemetry Transport (MQTT) protocol message queues are used to deliver IoT data streams at disparate
rates, which MDML then aggregates, preprocesses, and delivers to analysis tasks and ML models.
Any number of sensors or instrument groups can be created inside MDML. 
Data sets are treated as time series data and MDML's data fusion capabilities
enable diverse sources to be aggregated based on customizable batching rules.
Thus, MDML is a uniquely powerful environment for 
reinforcement learning and leveraging deep neural networks to sense key events
and steer experiments.

MDML is designed as a portable environment that can be configured to 
use a wide
variety of available computing resources, from edge to exascale, 
depending on computational demands and problem complexity.
MDML can also use commercial cloud environments such as AWS. 
To achieve this we have integrated the funcX~\cite{funcx} function serving platform into MDML by implementing
custom Node-RED pallets for funcX (and for Globus~\cite{chard2016globus} for data transfer) that can be used in MDML pipelines.
We extend the MDML client with a Globus Auth~\cite{GlobusAuth} native client to enable secure offloading of computational tasks to different computing resources. Globus Auth enables users to authenticate and to interact securely with MDML; MDML can then introspect the user's tokens and acquire new tokens to request actions (such as data transfers and funcX invocations) on
the user's behalf.

Instrument data are packaged in envelopes when streamed to MDML via MQTT topics.
These packages, which describe the experiment and data source, are used by MDML to route the data through the processing pipeline.
Once processed, data are compressed and archived in a persistent object store. The archived datasets contain descriptions of the streamed data to inform collaborators and enable post situ analyses.  

MDML provides dashboards as a gateway to data generated by an experiment and to support the building of analysis pipelines. MDML employs the industry standard Grafana dashboard as a customizable platform to create dynamic and interactive visualization experiences. MDML users develop their own dashboards to visualize sensor measurements, the state of analysis tasks, and results as data are processed. This not only allows researchers to dissect their data in real-time but also acts as a monitoring tool for critical experimental data.

\section{Demonstration}

Our demonstration will showcase MDML's application to a combustion experiment for high-throughput manufacturing of nanomaterials at Argonne's MERF. This experiment manufactures nanomaterials in high volumes using flame spray pyrolysis (FSP)~\cite{fsp}, a versatile process that allows for commodity-scale production of a broad range of nanomaterials. The FSP instrument includes multiple sensors, including a Planar Laser Induced Fluorescence (PLIF) diagnostic system that uses a tunable laser light sheet to characterize flame chemistries, spectroscopy to determine the contents of the exhaust and particle size distribution of the resulting nanomaterials. 
These sensors generate data at vastly different volumes and rates. For example, PLIF data can be generated every 50ms, whereas spectroscopy and particle size results are integrated every few minutes.

We will demonstrate the analysis pipeline used to perform near-real time quality control of the flame's stability. We will show how MDML enables scientists to both monitor the state of the experiment and steer the evolution of the flame. Using local resources for rapid quality control, HPC  resources for large-scale analysis, and ML models to integrate diverse data types, we will use MDML to guide flame stability. Finally, we will visualize experiments in real time via an interactive interface built on the industry standard Grafana platform.

\section*{Acknowledgement}
This work was supported by the U.S. Department of Energy, Office of Science, under contract DE-AC02-06CH11357.

\bibliographystyle{IEEEtran}
\bibliography{mdml}

\end{document}